\documentclass[preprintnumbers,amsmath,11pt,amssymb,floatfix,superscriptaddress,nofootinbib]{article}

\def\mytitle#1{\setcounter{equation}{0}
\setcounter{footnote}{0}
\begin{flushleft}\Large\textbf{#1}\end{flushleft}
\vspace{0.25cm}}
\def\myname#1{\leftline{{\large #1}}\vspace{-0.13cm}}
\def\myplace#1#2{\small\begin{flushleft}\textit{#1}\\
\texttt{#2}\end{flushleft}}

\def\myclassification#1{\small\noindent
Keywords :
       #1\vspace{0.5cm}}
\usepackage{graphicx}
\begin{document}
\mytitle{Thermodynamic studies  of Different Type of Black Holes: General Uncertainty Principle Approach}

\myname{$Amritendu~ Haldar^{*}$\footnote{amritendu.h@gmail.com} and $Ritabrata~
Biswas^{\dag}$\footnote{biswas.ritabrata@gmail.com}}
\myplace{* Department of Physics, Sripat Singh College, Jiaganj, Murshidabad $-$ 742123, India.\\$\dag$ Department of Mathematics, The University of Burdwan, Golapbag Academic Complex, City : Burdwan $-$ 713104, Dist. : Purba Burdwan, State : West Bengal , India.} {}
 
\begin{abstract}
We present an investigation on thermodynamics of two different types of black holes viz. Kiselev black hole (asymptotically flat) and Taub-NUT (non-asymptotically flat) black hole. We compute the thermodynamic variables like black hole's Hawking temperature, entropy at the black hole's event horizon.  Further we derive the heat capacity and examine it to study the thermal stability of the black holes. We also calculate the rate of emission assuming the black holes radiate energy in terms of photons by tunneling.  We represent all the parameters including the rate of emission of the black holes graphically and interpret them physically. We depict a comparative study of thermodynamics between the afroesaid types of black holes. Here we find the existence of a transition of phase. Finally, we obtain the quantum-corrected thermodynamics on the basis of general uncertainty principle and it is seen from the quantum-corrected entropy that it contains the logarithmic term. We offer comparative studies on joint effect of generalised uncertainty principle parameter $\alpha$ along with the concerned black holes' parameters on the thermodynamics.
\end{abstract}

\myclassification{Quintessence; quantum tunneling; Generalized Uncertainty Principle; quantum-corrected Entropy;  gravito-electric mass; gravito-magnetic mass.}

\section{Introduction}
Black holes (BHs hereafter) are the stellar end state compact objects. These might be treated as the solutions of the Einstein's general relativity where a central singularity, some inner horizons like Cauchy horizons etc are wrapped by an event horizon. Quantum fluid theoretical studies \cite{Bardeen 1973, Hawkings 1974} of such dense objects reveals that they even can radiate/ evaporate. This idea lead us to consider the BHs as thermodynamic systems. This particular type of radiations are named as Hawking radiation, Unruh radiation etc. In tunneling method, the particles such as photons may cross the event horizon of BHs by quantum tunneling procedure. Except this approach, one can use $(i)$ the Hamilton-Jacobi method \cite{Shankaranarayanan 2002, Agheben 2005, Kerner 2006, Akhmedov 2007, Akhmedov 2009} or can apply   $(ii)$ the radial null-geodesic method \cite{Parikh 2000, Jiang 2006, Xu 2007} to determine the tunneling result, i.e., Hawking temperature and entropy of the BHs. Considering self-gravitation and back reaction effect, the study of Hawking radiation is done using tunneling formalism  \cite{Banerjee 2008}. For self-dual BH, the back reaction effect is investigated in \cite{Silva 2013}. The information loss paradox in the WKB/ tunneling picture of Hawking radiation considering the back effects is computed in \cite{Singleton 2010, Singleton 2014}.

The tunneling method might be used as very much useful tool to calculate the quantum-corrected Hawking temperature and corresponding entropy. Entropy calculation incorporates the application of generalized uncertainty principle (GUP hereafter). The quantum-corrected Hawking temperature and entropy of a Schwarzschild BH with effects of GUP have been investigated in \cite{Majumder 2013}. There exists several works \cite{Casini 2011, Wang 2014, Magan 2014} in literature in which the authors proposed the way to understand the quantum aspects of the BH entropy. The quantum corrections to the Bekenstein-Hawking entropy is logarithamic and depend on area which haven focused by the authors in \cite{Kaul 2000}. The GUP has been applied to various BHs- self dual BHs \cite{Anacleto 2015}, two-dimentional Horava-Lifshitz BH \cite{Anacleto 2016}, brane world BHs \cite{Casadio 2017}, Bardeen regular BH \cite{Maluf 2018} etc. 

Almost twenty years have passed since the discovery of the late time cosmic acceleration of the universe, based on the works \cite{DE1, DE2}. Proposition of exotic matters, name of which were coined as dark energy or quintessence, is done. It is assumed that these kind of negative pressure exerting homogeneousfluids are distributed all over the universe. Effect of such quintessence matter on the static spherically symmetric BH  metric is counted by Kiselev \cite{Kiselev}.

Thermodynamics and phase transitions in rotating Kiselev black hole have been studied in \cite{1610.00376}. First order approximation of horizon is used to calculate thermodynamic features for all values of dark energy EoS-s. This work mainly focusses on the values of thermodynamic parameters like areas, entropies, horizon radii, surface gravity, surface temperatures etc for the event horizon as well as Cauchy horizon and found the thermodynamic products of those quantities. The types of phase transitions have been found as well. Different thermodynamic relations for Kiselev and dilaton black holes are studied in \cite{1508.04761}. This work also deals with the thermodynamic products and tests which property is global and which is not.

Previously, we observe some thermodynamic studies \cite{abhijit} of BHs, where it was concluded that if a BH is surrounded by quintessence, it is likely to have an unstable large black hole. More negativity of  the value of the quintessence EoS implied more thermodynamically unstable nature of the concerned BH. Works for modified gravity BH also reveals that it is a tendency for BHs to become unstable as we shift further from the Einstein's general relativistic model \cite{RB}. In the present work, we mainly study the thermodynamic parameters, i.e., the mass of the BHs,   Hawking temperature, surface gravity and also the thermal heat capacity at event horizon and the rate of emission of energy through event horizon. Then  we reexamine the same for different BHs with the GUP. Our first motivation is to analyse the joint effect of the quintessence and inclusion of GUP for a Kiselev BH). For even modified gravity theories, it was shown \cite{Dehghani} that NUT solutions are not thermally stable. Our second motivation for this letter is to check what will be the effect of inclusion of GUP on the unstable nature of Taub-NUT BH. Kiselev metric is a general kind of spherically symmetric black hole solution which can take the form of a Schwarzschild, Reissner-Nordstrom etc depending on different values of $c$ and $\omega_q$. On the other hand Taub NUT is of non asymptotically flat nature which can be treated as a twisted one in four dimensions whose exterior space time rotates in a peculiar manner. Though the space time  has no curvature singularity, there exists a conical singularity along the axis of its rotation \cite{Ong}.Inside the BH horizon, the conical singularity has an infinite amount of angle excess at the poles. On the contrary, outside the BH horizon, the ``angle excess"turns to be complex. The Gaussian curvature is discontinuous at the pole across the event horizon, but diverges as one approaches the horizon from either side($-\infty$ if approaches from the inside and $+\infty$ from the out side). These two kinds of BHs will cover a large area of BH models studied.

The letter is organized as follows. In section $2$ we introduce the basic equations which we use through out the letter. In section $3$ we investigate the thermodynamic properties like the mass of the BHs, surface gravity, Hawking temperature, the thermal heat capacity at event horizon and rate of emission of energy through event hotizon and observe their behavior graphically and present their physical interpretations for Kiselev BH Surrounded by the quintessence and Taub-NUT BH. In section $4$ we reestablish the thermodynamics under GUP for the same series of BHs. Finally, we conclude this letter in section $5$.     
\section{Basic Equations:}
\subsection{Derivation of BH Temperature:}
\textbf{(i) Hawking's Approach:}\\

A conserved quantity may be constructed by using the Killing vector as:
\begin{equation}\label{ah5.equn1}
2\kappa \xi^\nu = -\nabla^\nu(\xi^\mu \xi_\mu),
\end{equation}
where $\nabla^\nu$ is a covariant derivative and $\kappa$ is constant along $\xi$ orbit, i.e., Lie derivative of $\kappa$ along $\xi$ vanishes. That is,
\begin{equation}\label{ah5.equn2}
{\cal L_\xi} {\kappa}=0.
\end{equation}
Generally $\kappa$ is constant over the horizon of BHs, known as surface gravity.
  
Again the surface gravity  $ \kappa $ of a BH can be defined as the magnitude of the gradient of the norm of horizon generating Killing field $ \chi^a = \zeta^a + \Omega \Psi^a $, evaluated at the horizon, if we assume that the event horizon of a BH is a Killing horizon, i.e., that null horizon generators are 'orbits' of a Killing field. That is, at the horizon, we have 
\begin{equation}\label{ah5.equn3}
\kappa^2: = -( \nabla^a\vert \Psi\vert)(\nabla_a\vert \Psi\vert)~~.
\end{equation}
According to the classical BH thermodynamics \cite{Bekenstein 1973, Bekenstein 1974, Hawking 1975, Hawking 1976}, the surface gravity of BHs, $ \kappa $, is connected with the temperature $ T $ known as Hawking temperature, given by,
\begin{equation}\label{ah5.equn4}
T=\frac{\kappa}{2\pi}. 
\end{equation}
\textbf{(ii) Tunneling Approach:}\\

The another way to derive the BH temperature is the quantum tunneling approach. The quantum tunneling effect permits that the particles inside the BH cross the event horizon. Therefore the calculation of the tunneling probability of this process is possible, as indicated in \cite{Agheben 2005, Kerner 2006, Kerner 2008}. In such a process, we are interested in radial trajectories. We consider the metric of a BH in 2-dimension near the horizon as: 
\begin{equation}\label{ah5.equn5}
ds^2 = -f(r)dt^2 + \frac{dr^2}{f(r)}.
\end{equation} 
Thus, the problem is entirely solved in the $t-r$ plane. The Klein-Gordon equation for a scalar field $\phi$ with mass $m_\phi$ is given as:
\begin{equation}\label{ah5.equn6}
\hbar^{2} g^{\mu \nu}\nabla_\mu \nabla_\nu \phi-m_\phi^2 \phi=0, 
\end{equation} 
and with the aid of $equation$ (\ref{ah5.equn5}), is written as: 
\begin{equation}\label{ah5.equn7}
-\partial_t^2 \phi+ f(r)^2\partial_r^2\phi+\frac{1}{2}\partial_r f(r)^2\partial_r\phi-\frac{m_\phi^2}{\hbar} f(r)\phi=0.
\end{equation}   
By using the WKB method, one can obtain the solution of $equation$ (\ref{ah5.equn7}) as:
\begin{equation}\label{ah5.equn8}
\phi (t, r)=exp\left[-\frac{i}{\hbar} {\cal I}(t, r)\right].
\end{equation}
For the lowest order in $\hbar$, substituting $equation$ (\ref{ah5.equn8}) in $equation$ (\ref{ah5.equn7}) one can evaluate the Hamilton-Jacobi equation as: 
\begin{equation}\label{ah5.equn9}
(\partial_t {\cal I})^2-f(r)^2(\partial_r{\cal I})^2-m_\phi^2 f(r)=0,
\end{equation}
with the action that generates (\ref{ah5.equn7}) is given as:
\begin{equation}\label{ah5.equn10}
{\cal I}(t, r)=-Et+W(r),
\end{equation}
where $E$ represents the radiation energy and the explicit form for $W(r)$, the spatial part of that action is given as:
\begin{equation}\label{ah5.equn11}
W(r)_{\pm}=\pm \int \frac{dr}{f(r)}\sqrt{E^2-m_\phi^2f(r)}~.
\end{equation}
The functions $W(r)_{\pm}$ indicate the outgoing $(+)$ and ingoing $(-)$ solutions respectively. Classically, the outgoing solutions, i.e., the solutions which cross the event horizon or is moving away from $r_h$, is not allowed. Therefore, to calculate the Hawking radiation outside the event horizon we shall focus on $W_+(r)$.

With the approximation for the function $f(r)$ near the event horizon $r_h$, i.e., $f(r)=f(r_h)+ f^{'}(r_h)(r-r_h)+...$,  $equation$ (\ref{ah5.equn11}) assumes the simple form as: 
\begin{equation}\label{ah5.equn12}
W_+(r)=\int\frac{dr}{f^{'}(r_h)} \frac{\sqrt{E^2-m_\phi^2f^{'}(r_h)(r-r_h)}}{(r-r_h)}=\frac{2\pi i E}{f^{'}(r_h)},
\end{equation}
where `dash' stands for the first order derivative with respect to $r$. Therefore, the tunneling probability for a particle with energy $E$ is given by,
\begin{equation}\label{ah5.equn13}
\Gamma\simeq exp [-2 Im {\cal I}]=exp\left[-\frac{4\pi E}{f^{'}(r_h)}\right].
\end{equation}
Comparing this equation with the Boltzmann factor, $exp [-\frac{E}{T}]$, Hawking temperature derived by the tunneling method is written as:
\begin{equation}\label{ah5.equn14}
T=\frac{E}{2 Im {\cal I}}=\frac{f^{'}(r_h)}{4\pi}.
\end{equation}\\
\textbf{(iii) Thermodynamic Approach:}\\

Stationary BH solutions of Einstein Field Equation (EFE) relate the first law of thermodynamics as:
\begin{equation}\label{ah5.equn15}
dM=\kappa\frac{dA}{8\pi G} + \Omega dJ + \phi dQ .
\end{equation}
The BH area at event horizon is given by, 
\begin{equation}\label{ah5.equn16}
 A=\int ^{2\pi}_{0} \int ^{\pi}_{0} \sqrt{g_{\theta \theta}(r_h){g_{\phi\phi}(r_+)}} {d\theta d\phi} = 4\pi r_h^2 
\end{equation}
and it is related to the entropy $S$ of BHs as:
\begin{equation}\label{ah5.equn17}
S=\frac{A}{4}= \pi r_h^2 
\end{equation}
Again
\begin{equation}\label{ah5.equn18}
dH=dM=TdS+VdP ,
\end{equation}
where $M$ and $H$ are the mass and the total gravitational enthalpy of the BHs.   
The $equation$ (\ref{ah5.equn18}) gives,
\begin{equation}\label{ah5.equn19}
T=\left(\frac{\partial H}{\partial S}\right)_P =\left(\frac{\partial M}{\partial S}\right)_P. 
\end{equation}
It is obvious that the expressions of temperature obtained from $equations$ (\ref{ah5.equn4}), (\ref{ah5.equn14}) and (\ref{ah5.equn19}) are same.\\ 
\subsection{Thermal Heat Capacity or Specific Heat :}

The thermal heat capacity or plainly the specific heat at event horizon of a BH is given as:
\begin{equation}\label{ah5.equn20}
C_h=\left( \frac{\partial M}{\partial T_h}\right)= \left(\frac{\partial M}{\partial r_h}\right)\left( \frac{\partial T_h}{\partial r_h}\right)^{-1}=T_h\left(\frac{\partial S_h}{\partial T_h}\right)=T_h\left(\frac{\partial S_h}{\partial r_h}\right)\left( \frac{\partial T_h}{\partial r_h}\right)^{-1} 
\end{equation}\\
It is important to verify, whether the BH is thermodynamically stable or unstable by the following way,\\
(i) the BH is thermodynamically unstable if the specific is negative .\\
(ii) for thermodynamically stable BH, the specific heat will be positive .\\
(iii) there is also a critical case where the specific heat blows up and that signifies a second order phase transition for such BHs.
\subsection{Generalized Uncertainty Principle:}
The Heisenberg's uncertainty principle may be generalized in the microphysics regime as \cite{Zhao 2003, Sun 2004, Kim 2007, Yoon 2007, Nouicer 2007, Anacieto 2014, Anacieto 2015, Anacieto 2016}:
\begin{equation}\label{ah5.equn21}
\triangle x \triangle p\geq \hbar\left(1-\frac{\alpha l_p}{\hbar}\triangle p+\frac{\alpha^2 l_p^2}{\hbar^2} \triangle p^2\right),
\end{equation}
where $l_p$ is the Planck's length of the order of $10^{-35}m$ and $\alpha$ is a dimensionless positive parameter. The $equation$ (\ref{ah5.equn21}) is known as general uncertainty principle (GUP hereafter). The quadratic form of GUP is given by,
\begin{equation}\label{ah5.equn22}
\triangle x \triangle p\geq \hbar\left(1+\frac{\alpha^2 l_p^2}{\hbar^2} \triangle p^2\right).
\end{equation}
The $eqn$ (\ref{ah5.equn21}) also can be rewritten as:
\begin{equation}\label{ah5.equn23}
\triangle p \geq \frac{\hbar \triangle x}{2\alpha^2 l_p^2}\left(1-\sqrt{1-\frac{4\alpha^2 l_p^2}{\triangle x^2}} \right).
\end{equation}
Taking $\frac{l_p}{\triangle x}\ll 1$, $\hbar=1$ and by applying Taylor series expansion, we have from the $equation$ (\ref{ah5.equn23}), the following expression : 
\begin{equation}\label{ah5.equn24}
\triangle p \geq \frac{1}{\triangle x}\left(1+\frac{2\alpha^2 l_p^2}{\triangle x^2}+... \right).
\end{equation}
Applying the saturated form of the uncertainty principle, viz. $E\triangle x\geq 1$, the $equation$ (\ref{ah5.equn24}) can be written as:
\begin{equation}\label{ah5.equn25}
E_G\geq E\left(1+\frac{2\alpha^2 l_p^2}{\triangle x^2}+... \right).
\end{equation} 
Therefore, the tunneling probability for a particle with corrected energy $E_G$ crossing the event horizon will be,
\begin{equation}\label{ah5.equn26}
\Gamma\simeq exp [-2 Im {\cal I_G}]=exp\left\{-\frac{4\pi E_G}{f^{'}(r_h)}\right\}.
\end{equation}
Comparing $equation$ (\ref{ah5.equn26}) with the Boltzmann factor $exp \{-\frac{E}{T}\}$, the quantum-corrected Hawking temperature becomes,
\begin{equation}\label{ah5.equn27}
T_G=T\left(1+\frac{2\alpha^2 l_p^2}{\triangle x^2}+... \right)^{-1}.
\end{equation}\\
\section{Thermodynamic Properties:}
\subsection{Kiselev BH Surrounded by the Quintessence:}
In presence of quintessence matter, the metric of  Kiselev BH \cite{Kiselev 2003, Mubasher 2015} can be expressed as:
\begin{equation}\label{ah5.equn28}
ds^2 = -f(r)dt^2 + \frac{dr^2}{f(r)} + r^2 (d\theta^2 + \sin^2 \theta d\phi^2),
\end{equation}
where
\begin{equation}\label{ah5.equn29}
f(r)= 1-\frac{2M}{r}- \frac{c}{r^{3\omega_q+1}},
\end{equation}
where $M$ is the mass of BH, $c$ is the quintessence parameter and $\omega_q$ is the equation of state of exotic matter which is ranging from $-1$ to $-\frac{1}{3}$. Here we consider $\omega_q=-\frac{2}{3}$.

For vanishing $c$, the lapse function resembles to the Schwarzschild metric. With slight increase of $c$, $f(r)$ decreases slightly but due to increase of $r$, initially $f(r)$ increases rapidly and reaches a maximum. Further for increment of $r$, $f(r)$ decreases but with low rate. It is evident from $equation$ (\ref{ah5.equn29}) that there is a curvature singularity at $r=0$. $f(r)=0$ gives Killing horizons or BH horizons as:
\begin{equation}\label{ah5.equn30}
r_{h/C}=\frac{1}{2c}\left(1\pm \sqrt{1-8Mc}\right),
\end{equation} 
where $r_h$ and $r_C$ known as event horizon and Cauchy horizon respectively. It is clear from $equation$ (\ref{ah5.equn30}) that the horizons exist only when $ M<\frac{1}{8c}$.

For event horizon, the mass of the BH is expressed as:
\begin{equation}\label{ah5.equn31}
M=\frac{r_h}{2}(1-c r_h).
\end{equation} 

Using $equation$ (\ref{ah5.equn14}) one can compute the temperature of the BH at event horizon as:
\begin{equation}\label{ah5.equn32}
T_h=\frac{1}{4\pi r_h}\left(1-2cr_h\right).
\end{equation}
Equating this with the $equation$ (\ref{ah5.equn4}) we have the surface gravity of the BH at the event horizon as: 
\begin{equation}\label{ah5.equn33}
\kappa_h=\frac{1}{2 r_h}\left(1-2cr_h\right).
\end{equation}
Again equating $equation$ (\ref{ah5.equn32}) with $equation$ (\ref{ah5.equn19}) and using $equation$ (\ref{ah5.equn31}) we obtain the entropy of the BH at event horizon as:
\begin{equation}\label{ah5.equn34}
S_h=\pi r_h^2,
\end{equation}
as expected. This is actually the zeroth entropy $S_0$ \cite{XU 2015}. It is also evident that this expression is the direct proof of the area and the entropy relation of BHs as shown in $equation$ (\ref{ah5.equn17}).

The specific heat of the BH is computed by using $equations$ (\ref{ah5.equn31}) and (\ref{ah5.equn32}) in $equation$ (\ref{ah5.equn20}) as:
\begin{equation}\label{ah5.equn35}
C_h=-2\pi r_h(1-2cr_h).
\end{equation}
Therefore, it is obvious that the BH is thermodynamically stable under the condition that $r_h>\frac{1}{2c}$.

Assuming that the BH radiates the energy in terms of photon \cite{Emparan 2000, Tawfik 2013} and using the Stefan-Boltzmann law in 2-D space time we compute the rate of emission as:
\begin{equation}\label{ah5.equn36}
\frac{dM}{dt}\propto T_h^2,
\end{equation}
where `$t$' stands for the time between which the emission occurred.\\
Thus, in this case the rate of emission will be,
\begin{equation}\label{ah5.equn37}
\frac{dM}{dt}\propto \frac{1}{16\pi^2 r_h^2}\left(1+4c^2 r_h^2-4cr_h\right).
\end{equation}
\begin{figure}[h!]
\begin{center}
\includegraphics[scale=.8]{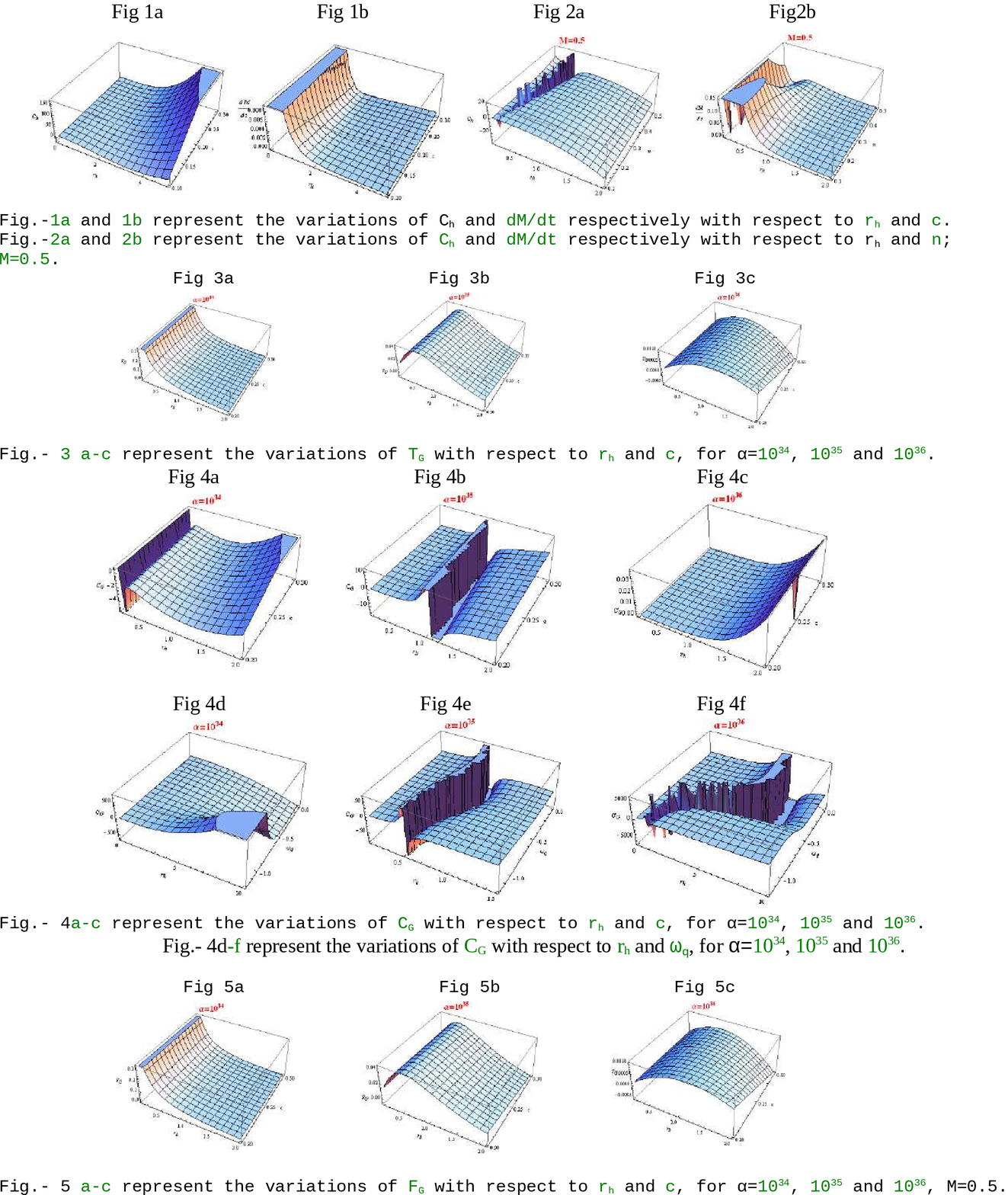}
\end{center}
\end{figure}
The variation of $ C_h $  with respect to $ r_h $ and $c$ is depicted in Fig.-1a for $equation$ (\ref{ah5.equn35}). Here we observe that $C_h$ is increased due to increase of $r_h$, though the rate of this increase is very slow for low values of $r_h$. $C_h$ increases with high rate and attains a maximum due to slow increment of $c$.

In Fig.-1b, we have plotted the variation of the rate of emission $ \frac{dM}{dt} $  with respect to $ r_h $ and $c$ for $equation$ (\ref{ah5.equn37}). It is seen from the curve that $ \frac{dM}{dt} $ decreases steeply with increase of $r_h$ at first, then the rate of decrease becomes low with further increment of $r_h$ keeping $c$ fixed. For increase of $c$, the general nature of $ \frac{dM}{dt} $ remains almost unchanged. Only the rate of decreadse for low $r_h$ is increased for high $c$.

\subsection{Taub-NUT BH}
The metric of  Taub-NUT (Newman, Unti and Tamburino)  BH \cite{Taub 1951, Misner 1963, Miller1 1971, Miller2 1973} is given as:
\begin{equation}\label{ah5.equn38}
ds^2 = -f(r)(dt+2n cos\theta d\phi)^2 + \frac{dr^2}{f(r)} + (r^2+n^2) (d\theta^2 + \sin^2 \theta d\phi^2),
\end{equation}
where
\begin{equation}\label{ah5.equn39}
f(r)= 1-\frac{2(Mr+n^2)}{r^2+n^2},
\end{equation}
where $M$ represents ADM mass or the gravito-electric mass and $n$ indicates the dual mass or gravito-magnetic mass of the BH. Actually, The Taub-NUT space-time \cite{Newman 1963, Taub 1951} is a stationary, spherically symmetric and non-asymptotically flat solution of the vacuum Einstein  field equation in
general relativity (GR hereafter).
When $n$ is very low, the metric given by equation (\ref{ah5.equn38}) and equation (\ref{ah5.equn39})is similar to the Schwarzschild metric. With slight increase of $r$, $f(r)$ increases rapidly at first and then the rate of increment slows down (but this variation does not depend upon $n$). 

It is evident from $equation$ (\ref{ah5.equn39}) that there is a curvature singularity at $r=0$, $f(r)=0$ gives Killing horizons or BH horizons as:
\begin{equation}\label{ah5.equn40}
r_{h/C}=M \pm \sqrt{M^2+n^2},
\end{equation} 
where $r_h$ and $r_C$ denote event horizon and Cauchy horizon respectively.

The mass of this BH at the  event horizon is computed as:
\begin{equation}\label{ah5.equn41}
M=\frac{r_h^2-n^2}{2r_h}.
\end{equation} 
We notice that $M$ increases linearly due to increment of $r_h$ keeping $n$
fixed. For low values of $r_h$ and high values of $n$ firstly $M$ increases rapidly. Latter its increment slows down.  

Applying the  $equation$ (\ref{ah5.equn14}), we can obtain the temperature of this BH at event horizon as:
\begin{equation}\label{ah5.equn42}
T_h=\frac{1}{2\pi}\frac{M(r_h^2-n^2)+2nr_h^2}{(r_h^2+n^2)^2}.
\end{equation}
Also equating this with the $equation$ (\ref{ah5.equn4}) we calculate the surface gravity of this BH at the event horizon as: 
\begin{equation}\label{ah5.equn43}
\kappa_h=\frac{M(r_h^2-n^2)+2nr_h^2}{(r_h^2+n^2)^2}.
\end{equation}
Equating $equation$ (\ref{ah5.equn42}) with $equation$ (\ref{ah5.equn19}) and applying $equation$ (\ref{ah5.equn41}) we get the entropy of this BH at event horizon as:
$$S_h=\frac{\pi}{3M^4}\Bigg\{3M^3n^4+12Mn^2(M^2+n^2)r_h-3M^2n^2r_h^2+M^3r_h^3-24n^3(M^2+n^2)^{\frac{3}{2}}Arctanh\left(\frac{n^2+Mr_h}{n\sqrt{M^2+n^2}}\right)$$
\begin{equation}\label{ah5.equn44}
-5M^2n^4 ln r_h-12n^4(M^2+n^2)ln (Mr_h^2+2n^2-Mn^2)\Bigg\}.
\end{equation}
 
The specific heat of the BH is obtained by using $equations$ (\ref{ah5.equn41}) and (\ref{ah5.equn42}) in $equation$ (\ref{ah5.equn20}) as:
\begin{equation}\label{ah5.equn45}
C_h=\frac{\pi(r_h^2+n^2)^4}{2r_h^2(n^4-Mr_h^3-3n^2r_h^2+3Mn^2r_h)}.
\end{equation}
For  this BH, the rate of emission is given as:,
\begin{equation}\label{ah5.equn46}
\frac{dM}{dt}\propto \frac{1}{4\pi^2}\frac{\bigg\{M(r_h^2-n^2)+2n^2r_h\bigg\}^2}{(r_h^2+n^2)^4}.
\end{equation}
In Fig.-2a, we have plotted the variation of $ C_h $ with respect to $ r_h $ and $n$ for $equation$ (\ref{ah5.equn45}). Here we observe a sharp peak for low values of $r_h$. This means there a phase transition occurs which is not shown in Kiselev BH. For increment of $r_h$, $ C_h $ decreases but the change in $n$ does not effect $C_h$. Then the BH is thermodynamically stable only when $\pi(r^2+n^2)^4>2r^2(n^4-Mr^3-3n^2r^2+3Mn^2r)$. 

We have plotted $\frac{dM}{dt}$ with respect to $r_h$ and $n$ in Fig.-2b for $equation$ (\ref{ah5.equn46}). It is observed that if $n$ is small, $\frac{dM}{dt}$ is a decreasing function of $r_h$. When $r_h$ is low, the rate of this decrease is high and latter the rate decreases and makes the curve asymptotic to a $r_h$ constant plane. As we increase $n$, after a certain value of $n$, we observe that the $\frac{dM}{dt}$ decreases first. Then it again increases and then again decreases, firstly with a higher slope and then with a lower slope.
\section{Thermodynamics under General Uncertainty Principle:}
\subsection{Kiselev BH Surrounded by the Quintessence:}
Here we assume that the uncertainty in $x$ for events near the event horizon is given by $\triangle x\simeq 2r_h$ \cite{Medved 2004}. The quantum-corrected temperature of the BH is thus,  
\begin{equation}\label{ah5.equn47}
T_G=T_h\left(1+\frac{\alpha^2 l_p^2}{2r_h^2}+... \right)^{-1}=T_h\left(1+\frac{2\alpha^2 l_p^2 c^2}{(1+\sqrt{1-8Mc})^2}+... \right)^{-1}.
\end{equation}
We have plotted the variations of GUP corrected Hawking temperature, $T_G$ with respect to $r_h$ and Kiselev parameter $c$ for $\alpha=10^{34}$, $10^{35}$ and $10^{36}$ in Fig.-3a-c respectively. For $\alpha=10^{34}$, we observe that $T_G$ is a decreasing function of $r_h$ and this nature is not effected much for the variation of $c$. But as we increase $\alpha$ to $10^{35}$, the scenario changes. In general, $T_G$ increases and then after reaching to a maxima at a certain $r_h=r_{h-crit_1}(c,\alpha)$ it starts to decrease with $r_h$. If $c_1>c_2$, we see $r_{h-crit_1}(c_1, \alpha)<r_{h-crit_1}(c_2, \alpha)$. This means if $c$ is high, the maxima is taking place earlier or a smaller BH gets hotter in more early age/size. Physically saying, more effective is the quintessence, the BH . For $\alpha=10^{36}$ case, the steepness of increment and decrease(after reaching the maxima) both slows down. Mathematically speaking, if $\alpha_1>\alpha_2$ then $r_{h-crit_1}(c, \alpha_1)>r_{h-crit_1}(c, \alpha_2)$.

Now, the quantum-corrected entropy is computed as 
\begin{equation}\label{ah5.equn48}
S_G=S_h-\pi \alpha^2 l_p^2 lnr_h+....
\end{equation}
The quantum-corrected entropy term contains the logarithmic term of event horizon $r_h$ as shown in \cite{Kaul 2000}.

If we apply quantum-corrected temperature and quantum-corrected entropy, we can calculate the quantum-corrected heat capacity by the relation,
\begin{equation}\label{ah5.equn49}
C_G=\left( \frac{\partial M}{\partial T_G}\right)= \left(\frac{\partial M}{\partial r_h}\right)\left( \frac{\partial T_G}{\partial r_h}\right)^{-1}=T_G\left(\frac{\partial S_G}{\partial T_G}\right)=T_G\left(\frac{\partial S_G}{\partial r_h}\right)\left( \frac{\partial T_G}{\partial r_h}\right)^{-1} 
\end{equation}
and that can be expressed as:
\begin{equation}\label{ah5.equn50}
C_G=\frac{C_h\left(1+\frac{\alpha^2 l_p^2}{2r_h^2}+... \right)}{1-(1-2cr_h)\left(1+\frac{\alpha^2 l_p^2}{2r_h^2}+... \right)^2\left(\frac{\alpha^2 l_p^2}{r_h^2}+...\right)}
\end{equation}
Fig.-4a-c depict the variation of the GUP supported specific heat with $r_h$ and $c$. For $\alpha=10^{34}$, we see the specific heat is negative for low $r_h$. As $r_h$ increases, we observe the specific heat decreases firstly and then achieving a minima, it again increases and then it becomes positive for some $r_h=r_{h-crit_2}(c,\alpha)$. For $c_1>c_2$, we see $r_{h-crit_2}(c_2,\alpha)<r_h=r_{h-crit_2}(c_1,\alpha)$. Physically, this says that the BH is unstable if it is small. As the BH grows bigger it transits through a first order phase transition and it becomes stable large BH. Increment in $c$ inhibits the BH to quickly transit. For $\alpha=10^{35}$, the BH is stable and small firstly. Then it passes through a second order phase transition and latter it becomes unstable and bigger one. Here also $c$ plays a crucial role. If $c$ increases, the second order phase transition occurs early or for smaller values of $r_h$ . $\alpha=10^{36}$ case shows stable BH firstly. The phase transition occurs lately.

Fig.-4d-f represent the variation of $C_G$ with respect to the radius of event horizon $r_h$ and quintessence EoS, $\omega_q$, for $\alpha=10^{34}$, $\alpha=10^{35}$ and  $\alpha=10^{36}$ respectively. We will study $\alpha=10^{34}$ case at first. When $\omega_q =0$, i.e., pressureless dust is concerned, we see the specific heat is negative and decreasing. This says if a Schwarzschild BH is concerned, the BH is unstable. This is a known result however. But as we decrease $\omega_q$, i.e., the negative pressure exerting switch is on, we observe a positive specific heat depicting the fact that a stable small BH can be found when it is embedded in dark energy universe. Now, as $\alpha$ is increased to $10^{35}$, the existence of stable small BH followed by a second order phase transition and formation of a larger unstable BH are pointed. If $\omega_q$ is low, i.e., dark energy is more effective, the phase transition arises for lower value of radius of event horizon. So, effect of dark energy around a BH does not allow the presence of stable BH. When $\alpha$ is equal to $10^{36}$, this incident becomes more prominent. So we can conclude that the effect of quintessence is increased as we take the GUP's effect more prominently. GUP inhibits the effect of quintessence.

It is obvious that for $\alpha\longrightarrow 0$, we obtain the heat capacity as $equation$ (\ref{ah5.equn35}), i.e.,
\begin{equation}\label{ah5.equn51}
lim_{\alpha\longrightarrow0}C_G=C_h
\end{equation}
The quantum-corrected Helmholtz free energy is expressed as:
\begin{equation}\label{ah5.equn52}
F_G=M-T_GS_G
\end{equation}
Here we apply the classical BH thermodynamics \cite{Bekenstein 1973, Bekenstein 1974, Hawking 1976} prediction, i.e.,  $M$ = $U$ the internal energy of the BH. Thus
\begin{equation}\label{ah5.equn53}
F_G=M-T_h\left(1+\frac{\alpha^2 l_p^2}{2r_h^2}+... \right)^{-1}\left(S_h-\pi \alpha^2 l_p^2 lnr_h+....\right)
\end{equation}
 
Fig.-5a-c are the descriptions of the variations of free energy with respect to $r_h$ and $c$. For $\alpha=10^{34}$, if $c$ is low, free energy is decreasing. When $c$ is increased, free energy is firstly decreasing and then increasing. If $c$ is high free energy is only increasing. For $\alpha=10^{35}$ case, if $c$ is low, $F_G$ increases first, then decreases and then again increases. The region of first increment and decrease  will be shortened for high $c$. For $\alpha=10^{36}$, this incrase and decrease slows down.

The quantum-corrected rate of emission is given as:
\begin{equation}\label{ah5.equn54}
\frac{dM}{dt}\propto T_G^2
\end{equation}
and which gives,

\begin{equation}\label{ah5.equn55}
\frac{dM}{dt}\propto \frac{1}{16\pi^2 r_h^2}\left(1+4c^2 r_h^2-4cr_h\right)\left(1+\frac{\alpha^2 l_p^2}{2r_h^2}+... \right)^{-2}.
\end{equation}
\subsection{Taub-NUT BH:}
Due to the same assumption as previous, we have the quantum-corrected temperature of this BH which is given by,  
\begin{equation}\label{ah5.equn56}
T_G=T_h\left(1+\frac{\alpha^2 l_p^2}{2r_h^2}+... \right)^{-1}=T_h\left(1+\frac{\alpha^2 l_p^2 }{2(M+\sqrt{M^2+n^2})^2}+... \right)^{-1}.
\end{equation}
Fig.-6a-c are for Haking temperature. We count the variations with respect to $r_h$ and $n$ for $\alpha=10^{34}$, if $n$ is low, we observe that the temperature is decreasing. For high value of $n$, Hawking temperature is firstly increasing, then almost constant. If we increase $\alpha$ to $10^{35}$ or $10^{36}$, we see the temperature to increase for ever.

The quantum-corrected entropy for $Taub-NUT$ BH is expressed as:
$$S_G=\frac{\pi}{6M^4}\Bigg\{2M^3 r_h^3-6M^2n^2r_h^2+l_p^2M^3n^4\alpha^2r_h^{-3}+3l_p^2M^2n^4\alpha^2r_h^{-2}+3Mr_h\{8n^4+M^2(8n^2+l_p^2 \alpha^2)\}$$
$$+6Mn^2\{2l_p^2n^2\alpha^2+M^2(n^2+2l_p^2\alpha^2)\}r_h^{-1}-24n(M^2+n^2)^{\frac{3}{2}}(2n^2+l_p^2\alpha^2)Arctanh\left(\frac{n^2+Mr_h}{n\sqrt{M^2+n^2}}\right)$$
 \begin{equation}\label{ah5.equn57}
-6\{4l_p^2n^4\alpha^2+M^2(2n^4+5l_p^2n^2\alpha^2)\}ln r_h-12n^2(M^2+n^2)(2n^2-l_p^2\alpha^2)ln(Mr_h^2+2n^2r_h-Mn^2)\Bigg\}
\end{equation}
The quantum-corrected entropy term also contains the logarithmic term of event horizon $r_h$ as shown in \cite{Kaul 2000}.\\
Using quantum-corrected temperature and quantum-corrected entropy, one can calculate the quantum-corrected heat capacity given as:,
\begin{equation}\label{ah5.equn58}
C_G=\left( \frac{\partial M}{\partial T_G}\right)= \left(\frac{\partial M}{\partial r_h}\right)\left( \frac{\partial T_G}{\partial r_h}\right)^{-1}=T_G\left(\frac{\partial S_G}{\partial T_G}\right)=T_G\left(\frac{\partial S_G}{\partial r_h}\right)\left( \frac{\partial T_G}{\partial r_h}\right)^{-1} 
\end{equation}\\
and that can be expressed as:
\begin{equation}\label{ah5.equn59}
C_G=\frac{4\pi r_h^3\left(r_h^2+n^2\right)^4\left(1+\frac{\alpha^2 l_p^2}{2r_h^2}+...\right)}{2r_h^2\{4r_h^3(n^4-Mr_h^3-3n^2r_h^2+3Mn^2r_h)-n^2\alpha^2 l_p^2(2r_h^3-6Mr_h^2-6n^2r_h+2Mn^2)\}}
\end{equation}

Fig.-7a-c are for specific heats. The general view is : firstly we see a stable phase from where the BH transits to an unstable phase via a second order phase transition. If the phase transition occurs at $r_h=r_{h-crit_3}(n,\alpha)$(say), then we see  $r_{h-crit_3}(n_1,\alpha)> r_{h-crit_3}(n_2,\alpha)$ for $n_1>n_2$. Again we observe $r_{h-crit_3}(n,\alpha_1)> r_{h-crit_3}(n,\alpha_2)$ for $\alpha_1>\alpha_2$. Free energy's curves are given in Fig.-8a-c. The curves are increasing then decreasing for $r_h$.

From the above equation it is found that, for $\alpha\longrightarrow 0$, we have the heat capacity as $equation$ (\ref{ah5.equn45}), i.e.,
\begin{equation}\label{ah5.equn60}
lim_{\alpha\longrightarrow0}C_G=C_h
\end{equation}
The quantum-corrected Helmholtz free energy for this BH is derived as:
$$F_G=M-T_h\left(1+\frac{\alpha^2 l_p^2}{2r_h^2}+... \right)^{-1}\frac{\pi}{6M^4}\Bigg\{2M^3 r_h^3-6M^2n^2r_h^2+l_p^2M^3n^4\alpha^2r_h^{-3}+3l_p^2M^2n^4\alpha^2r_h^{-2}+3Mr_h\{8n^4+M^2(8n^2+l_p^2 \alpha^2)\}$$
$$+6Mn^2\{2l_p^2n^2\alpha^2+M^2(n^2+2l_p^2\alpha^2)\}r_h^{-1}-24n(M^2+n^2)^{\frac{3}{2}}(2n^2+l_p^2\alpha^2)Arctanh\left(\frac{n^2+Mr_h}{n\sqrt{M^2+n^2}}\right)$$
 \begin{equation}\label{ah5.equn61}
-6\{4l_p^2n^4\alpha^2+M^2(2n^4+5l_p^2n^2\alpha^2)\}ln r_h-12n^2(M^2+n^2)(2n^2-l_p^2\alpha^2)ln(Mr_h^2+2n^2r_h-Mn^2)\Bigg\}
\end{equation}
\begin{figure}[h!]
\begin{center}
\includegraphics[scale=.8]{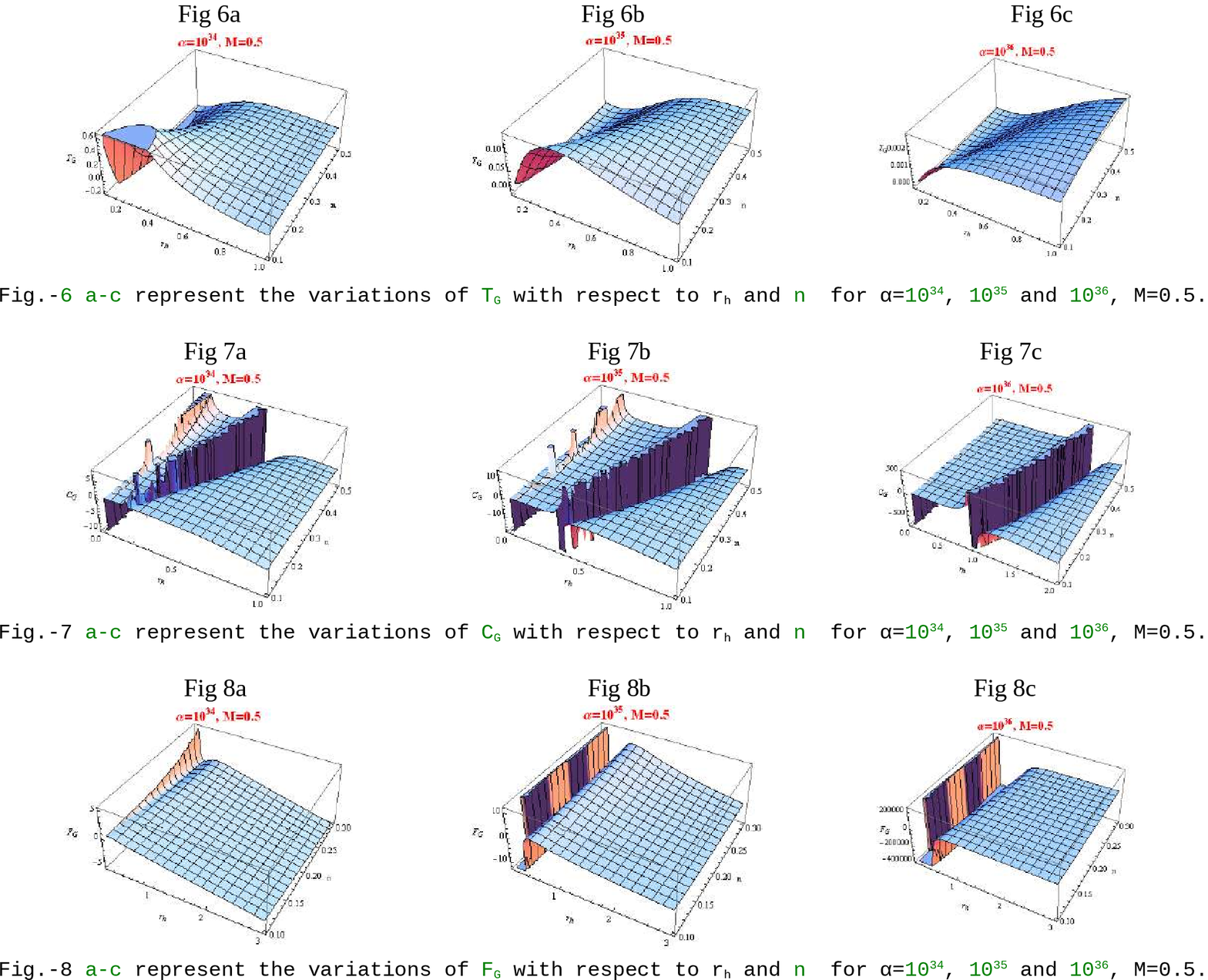}
\end{center}
\end{figure}
The quantum-corrected rate of emission is given as:
\begin{equation}\label{ah5.equn62}
\frac{dM}{dt}\propto T_G^2
\end{equation}
and which gives,
\begin{equation}\label{ah5.equn63}
\frac{dM}{dt}\propto \left\{\frac{M(r_h^2-n^2)+2nr_h^2}{2\pi(r_h^2+n^2)^2}\right\}^2\left(1+\frac{\alpha^2 l_p^2}{2r_h^2}+... \right)^{-2}.
\end{equation}
\section{Conclusion}
In this letter, we have investigated the thermodynamic parameters, i.e., the mass of the BHs and Hawking temperature, surface gravity at event horizon of  different BHs such as Kiselev BH Surrounded by the Quintessence and Taub-NUT BH which incorporates the magnetic field's effect in its lapse function. Moreover, we reexamine the Hawking temperature, heat capacity and the rate of emission on the basis of GUP for the same BHs. We have followed that for both type of BHs the quantum-corrected entropy term contains the logarithmic term of event horizon $r_h$ that is already proposed in \cite{Kaul 2000}. Regarding BH thermodynamics of simplest black hole model, Schwarzschild metric, it is considered that if the BH is small, its temperature will be large and so it will have a tendency to radiate out. Different modified gravity black holes show the same effect is mass is the only parameter inside them. When they are small, their temperature is high and as the radius of the event horizon grows bigger, the temperature decreases, first abruptly and latter slowly. A large black hole will have very low temperature and hardly radiates. In literature, we can find many studies of modified gravity black holes which keep some other thermodynamic parameters like charge, concerned modified gravity's coupling parameters other than the mass. A general tendency of the Hawking temperature for these kind of black holes is seemed to be increasing first and then after reaching a local maxima it starts to decrease. Now, the occurrance of maxima signifies the first order derivative of the temperature to be zero which forces the denominator of the expression of specific heat to vanish. This again creates an infinite jump discontinuity in the variation of the specific heat and we call it a second order phase transition. Question is : how large the black hole is when a second order phase transition may occur. In another way this can be stated as how long the stable regime of a black hole can last.

In this letter, when generalised uncertainty principle is incorporated, we observe the Hawking temperature of the Kiselev black hole to have the maxima. As we increased the effect of quintessence in it, we found even at a large radius of event horizon , the maxima is taking place. action of both the GUP and exotic matter forces the black hole to have the maximum temperature at higher radius of event horizon than the case when GUP is not properly inserted or effect of quintessence is not counted. These variations are given in Fig.-3a-c.

For Taub-NUT(Fig.-6a-c), if $n$, the gravito magnetic mass is nearly about 0.1, we see the Hawking temperature to decrease only when the GUP is taken to be low.But as we increase the value of $n$, for even the small effect of GUP, we find an occurrance of maxima. But as GUP's effect is increased by increasing the value of $\alpha$, we observe even for small value of $n$ the maxima takes place. For high $\alpha$ and high $n$, the maxima takes place for higher value of the radius of event horizon, $r_h$. So upto a huge size, the black holes temperature increases if we apply GUP properly and incorporate high magnetic mass. Magnetism, upto a high value of gravitating mass, is able to hold the central inward gravitating force.

Next, we will analyse the results collected from the specific heat curves. Before GUP's action, we see an unstable black hole is becoming stable via a first order phase transition for Kiselev black hole(Fig.-4a-c). More the effect of quintessence, more quick occurrance of the first order phase transition is observed. Inclusion of GUP changes the scenario. A second order phase transition transits a stable black hole to its unstable counterpart. Exotic matter's effect accelerates the phase transition to occur at smaller value of radius of event horizon. For $\alpha=10^{36}$, the second order phase transition occurs lately, i.e., for higher value of $r_h$. The phase transition occurs at larger radius of event horizon if GUP is acting but there is no quintessence(i.e., when $\omega_q=0$; pressureless dust). Once quintessence starts to work, this phase transition occurs earlier and at phantom era it is seen to happen so early. So we can conclude that the quintessence EoS parameter $\omega_q$ tries to form a second order phase transition. Increment in $c$ acts as (Fig.-4f) a catalyst and introduction of GUP works like an inhibitor in this phase transition process. For Taub-NUT case, the same pattern is observed. But whether GUP is working or not, it hardly does matter for the occurance of the second order phase transition. More effective is GUP for Taub-NUT, the second order phase transitrion occurs more lately. same is the effect of gravito magentic mass $n$. 

In a nutshell, we can conclude that both the ``set of quintessence parameters : ($c, \omega_q$) for Kiselev Black hole" and the ``gravito magnetic mass $n$ for Taub NUT black hole", inhibited by introduction of GUP, causes the occurance of a second order phase transition. Only the difference is that the occurance of second order phase transition is must for Taub NUT along with a possibility of having a large stable black hole before it transits. Second order phase transition is not mandatory for all the value of GUP parameter when Kiselev black hole is considered and before the the transition it is likely to have a small stable black hole.

\vspace{.1 in}
{\bf Acknowledgment:}
This research is supported by the project grant of Goverment of West Bengal, Department of Higher Education, Science and Technology and Biotechnology (File no:- $ST/P/S\&T/16G-19/2017$). AH wishes to thank the Department of Mathematics, the University of Burdwan for the research facilities provided during the work. RB thanks IUCAA, PUNE for Visiting Associateship.

\end{document}